%% file: main.tex
\documentclass{sig-alternate-10pt}
\usepackage[format=plain,labelfont=bf,up,textfont=it,up,small]{caption}

\usepackage{graphicx}
\usepackage{cite}
\usepackage[hyphens]{url}
\usepackage{float}
\usepackage{soul}
\usepackage{hyperref}
\usepackage{multirow}

\usepackage{amssymb}% http://ctan.org/pkg/amssymb
\usepackage{pifont}% http://ctan.org/pkg/pifont
\newcommand{\cmark}{\ding{51}}%
\usepackage{etoolbox}
\newcommand{\xmark}{\ding{55}}%

\usepackage{xspace}
\usepackage{tikz}
\usepackage{fontenc}
\newcommand*\circled[1]{\tikz[baseline=(char.base)]{
            \node[shape=circle,draw,inner sep=2pt] (char) {#1};}}

\makeatletter
\patchcmd{\maketitle}{\@copyrightspace}{}{}{}
\makeatother

\begin{document}

\title{Stress Testing the Booters: Understanding and Undermining the Business of DDoS Services}


\numberofauthors{3}
\author{
% 1st. author
\alignauthor
       Mohammad Karami \\
       \affaddr{George Mason University}\\
       %\affaddr{1932 Wallamaloo Lane}\\
       %\affaddr{Wallamaloo, New Zealand}\\
       %\email{mkarami@gmu.edu}
% 2nd. author
\alignauthor
       Youngsam Park\\
       \affaddr{University of Maryland, College Park}\\
       %\email{yspark@cs.umd.edu}
% 3rd. author
\alignauthor 
       Damon McCoy \\
       \affaddr{International Computer Science Institute}\\
       %\email{mccoy@cs.gmu.edu}
}

\maketitle% THIS IS SIGPROC-SP.TEX - VERSION 3.1
% WORKS WITH V3.2SP OF ACM_PROC_ARTICLE-SP.CLS
% APRIL 2009
%
% It is an example file showing how to use the 'acm_proc_article-sp.cls' V3.2SP
% LaTeX2e document class file for Conference Proceedings submissions.
% ----------------------------------------------------------------------------------------------------------------
% This .tex file (and associated .cls V3.2SP) *DOES NOT* produce:
%       1) The Permission Statement
%       2) The Conference (location) Info information
%       3) The Copyright Line with ACM data
%       4) Page numbering
% ---------------------------------------------------------------------------------------------------------------
% It is an example which *does* use the .bib file (from which the .bbl file
% is produced).
% REMEMBER HOWEVER: After having produced the .bbl file,
% and prior to final submission,
% you need to 'insert'  your .bbl file into your source .tex file so as to provide
% ONE 'self-contained' source file.
%
% Questions regarding SIGS should be sent to
% Adrienne Griscti ---> griscti@acm.org
%
% Questions/suggestions regarding the guidelines, .tex and .cls files, etc. to
% Gerald Murray ---> murray@hq.acm.org
%
% For tracking purposes - this is V3.1SP - APRIL 2009

\input{sections/abstract}

%\category{H.4}{Public Policy Issues}{ABUSE AND CRIME INVOLVING
%COMPUTERS}
%A category including the fourth, optional field follows...
%\category{D.2.8}{Software Engineering}{Metrics}[complexity measures, performance measures]


%\keywords{DDoS; Booter; Measurement} % NOT required for Proceedings

\input{sections/introduction}
\input{sections/background}
\input{sections/related}

\input{sections/users_victims}
\input{sections/infrastructure}
\input{sections/paypal_intervention}
\input{sections/discussion}
\input{sections/conclusion}
\input{sections/acknowledgments}

\begin{small}
\bibliographystyle{abbrv}
\bibliography{main}  % sigproc.bib is the name of the Bibliography in this case
\end{small}
\balancecolumns
% That's all folks!
\end{document}

%% file: sections/abstract.tex
\begin{abstract}
DDoS-for-hire services, also known as \textit{booters}, have commoditized DDoS attacks 
and enabled abusive subscribers of these services to cheaply 
extort, harass and intimidate businesses and people by knocking them offline. However, due
to the underground nature of these booters, little is known about 
their underlying technical and business structure.
In this paper we empirically measure many facets of their technical and payment 
infrastructure. We also perform an analysis of leaked and scraped
data from three major booters---Asylum Stresser, Lizard Stresser and VDO---which provides us
with an in-depth view of their customers and victims.
Finally, we conduct a large-scale payment intervention in collaboration with PayPal 
and evaluate its effectiveness.
Based on our analysis we show that these services are responsible for hundreds of thousands of 
DDoS attacks and identify potentially promising methods of 
increasing booters' costs and undermining these services.
\end{abstract}

%% file: sections/introduction.tex
\section{Introduction}

Distributed Denial-of-Service (DDoS) attacks are becoming a growing threat with
high profile DDoS attacks disrupting many large scale gaming services, such as
Microsoft's XBox Live and Sony's PlayStation networks at the end of 2014~\cite{xbox}.
These attacks were later claimed to be launched by the Lizard Squad as advertisements for 
their new DDoS-for-hire service called \textit{Lizard Stresser}~\cite{maketing}.
%As has been pointed out in previous studies, the volume of traffic focused at the victim
%is being increased (amplified) by orders of magnitude in some cases by abusing misconfigured servers 
%where the attacker sends small spoofed requests to these servers and far larger responses are sent from 
%these amplifiers back to the victim~\cite{ntp,ddoshell}. 
There is a long line of technical work exploring how to detect and mitigate these types of 
attacks~\cite{ddos_taxonomies,reflectors,Traceback,low_rate,pushback,botz,AmplificationHell,TCPAmplification}.

However, a large amount of DDoS attacks are being launched by relatively unsophisticated
attackers that have purchased subscriptions to low-cost DDoS-for-hire (commonly called booter) services. 
These services are operated by profit-motivated adversaries that have scaled up their DDoS infrastructure to
meet the increasing demand for DDoS attacks. Despite the threat they pose, little is known about the 
structures of these booter services and potential weaknesses in their operations that could be used to 
undermine them.

In this paper we undertake a large scale measurement study of
these booter services to understand how they are structured both technologically and economically 
with the focus of isolating potential weaknesses. We explore booters from
three different angles including analysis of leaked and scraped data, measurements of their
attack infrastructure and a payment intervention.

Our analysis of leaked and scraped
data from three booters---Asylum Stresser, Lizard Stresser and VDO~\footnote{We assign each booter service a unique three letter code based on their domain name to avoid
unintentionally advertising their services. The two exceptions are Asylum Stresser, which ceased operation before our study and Lizard Stresser,
which has already been highly publicized.}---demonstrates that these services 
have attracted over 6,000 subscribers and have launched over 600,000 attacks.
%Aggregate data provided by PayPal on the geolocation of subscribers' accounts suggests
%that the majority are located in the United States and Western Europe. 
We also find that the majority of booter customers prefer paying via PayPal
and that Lizard Stresser, which only accepted Bitcoin, had a minuscule 2\% sign-up to paid subscriber 
conversion rate compared to 15\% for Asylum Stresser and 23\% for VDO, which both accepted PayPal.
By analyzing attack traffic directed
at our own servers we are able to characterize the set of amplifiers they use to direct large amounts
of traffic at their victims. 
In order to measure the resilience of their payment infrastructure, we conduct a payment intervention 
in collaboration with PayPal. Our evaluation of the effectiveness
of this approach suggests that it is a promising method for reducing the subscriber base of booters.

In this paper, we further our understanding of the booter ecosystem through our measurements. 
Based on this we
identify potential improvements to ongoing efforts to disrupt their attack infrastructure
and an alternative and possibly more effective method of undermining 
these services by targeting their payment infrastructure. Overall, we find a 
few places where costs might be
marginally increased by more precisely mapping out and targeting parts of their attack infrastructure. We
document how a large-scale payment intervention by PayPal impacts booters, including
service closures. Finally, we detail some of their
strategies for evading detection by PayPal and discuss how these increase the
effort and costs associated with performing an ongoing payment intervention.

%% file: sections/background.tex
\section{Background}

In this section we explain the high level business and technical structure of booter services as well as the underlining ethical framework
for our measurements.

\subsection{Booter Services}

Thinly veiled booter services have existed since at least 2005 and primarily operate using
a subscription-based business model. 
As part of this subscription model, customers or subscribers~\footnote{We use these two terms interchangeably in this paper.}
can launch an unlimited number of attacks that have a duration typically ranging from 30 seconds to 1-3 hours and are limited
to 1-4 concurrent attacks depending on
the tier of subscription purchased. The price for a subscription normally ranges from \$10-\$300 USD per month depending on the
duration and number of concurrent attacks provided. 
These services claim that they are only to be used by network operators to 
stress test their infrastructure. However, they have become synonymous with DDoS-for-hire
and are a growing threat due to the fact that they have commoditized DDoS attacks that reach upwards of 2-3~Gbps. By offering a low-cost 
shared DDoS attack infrastructure,
these criminal support services have attracted thousands of malicious customers and are responsible for 
hundreds of thousands of DDoS attacks a year as we will show in Section~\ref{sec:scale}. 

These services can be found by visiting underground forums where they advertise and by web searches for terms, such as ``stresser'' 
and ``booter.''  
The services are all in English; we did not find any evidence of similar services that are focused on other markets, such as
Asian or Russian.
They maintain frontend sites that allow their customers to purchase subscriptions and launch attacks using
simple web forms. Their backend infrastructure commonly consists of databases that maintain 
subscriber information, lists of misconfigured hosts that can be used for DDoS amplification and most rent high-bandwidth Virtual 
Private Servers (VPS) for attacks rather than using botnets.
Ironically, booter services depend on DDoS-protection services, such as CloudFlare,
to protect their frontend and attack infrastructure from attacks launched by rival competing booter services. 

\begin{figure}[h]
  \centering
        \includegraphics[width=0.50\textwidth]{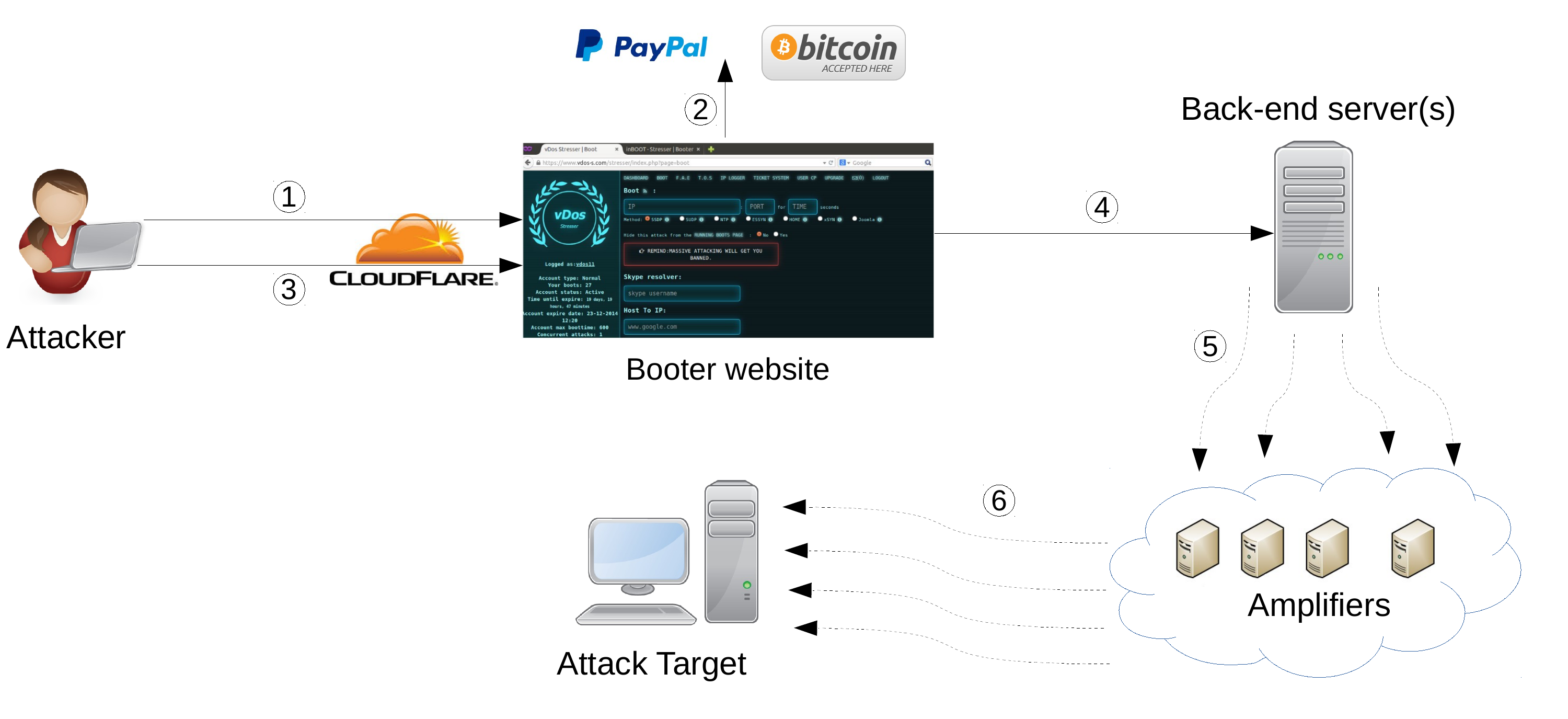}
        \caption{Structure of booter services.}
        \label{fig:booter_structure}
\end{figure}

Figure~\ref{fig:booter_structure}
provides a detailed illustration of the infrastructure and process of using a booter service. 
(\circled{1}) The customer first locates a booter site and visits their frontend webserver, 
which is normally protected by CloudFlare. (\circled{2}) The customer must next purchase a subscription using a payment method, such as
Bitcoin or PayPal. (\circled{3}) The customer then uses the frontend interface to request a DDoS attack against a victim. (\circled{4}) This request is forwarded
from the frontend server to one of the backend attack servers. (\circled{5}) The backend server then sends spoofed request packets to a set of
previously identified vulnerable amplification servers. (\circled{6}) Finally, DDoS traffic in the form of replies is sent to the victim from the abused
amplification servers.

\subsection{Ethical Framework}

As part of the ethical framework for our study, we placed a number of restrictions on the types of booter services we actively interacted with
and what we included in this paper. First, we did not engage with any DDoS service that advertised
using botnets. For example, in the case of Lizard 
Stresser, when we became aware that a botnet was being used, we immediately abandoned plans to collect active attack measurements from this service and restricted ourselves
to passive measurements. Our victim server was connected by a dedicated 1~Gbs network connection that was not shared with any other servers.
We also obtained consent from our network operators before conducting any DDoS attack experiments. When services used 
reflectors, we minimized the attack durations. 

When purchasing a subscription for a Booter service, we selected the cheapest option to
minimize the amount of money given to these services. In total, we spent less than \$140 and no individual booter service received more than \$19 in payments
as part of the measurements in this paper. Payments were made primarily using PayPal and we assumed that proper controls were put in place at
PayPal to mitigate the risk of money flowing to criminal groups. Also, the 9 booters that overlapped with our payment 
intervention study likely lost larger sums of money due to our reporting of their PayPal accounts than we paid to them.

We received an exemption from our Institutional Review Board (IRB), since our study did not include any personally identifiable information
and was based on publicly leaked data and measurements of services that are publicly accessible.
Finally, we did not disclose the identity of customers or operators of these services even when we became aware
of their identities and we did not reveal the identities of any victims unless they had been previously publicly disclosed.

%% file: sections/related.tex
\begin{table*}[ht!]
\small
\centering
%\resizebox{\linewidth}{!}{%
\begin{tabular}{lcrrrrr}
\hline
\bfseries Booter & \bfseries Period & \bfseries All Users & \bfseries Subscribers  & \bfseries Revenue & \bfseries Attacks & \bfseries Targets \\
\hline
Asylum Stresser & 10/2011-3/2013 & 26,075 & 3,963 & \$35,381.54 & 483,373 & 142,473 \\
Lizard Stresser & 12/2014-01/2015 & 12,935 & 176 & \$3,368 \(^\dagger\) & 15,998 & 3,907 \\
VDO\(^\ddagger\) & 12/2014-2/2015 & 11,975 & 2,779 & \$52,773* &138,010 & 38,539 \\
\hline
Total & - & 50,985 & 6,918 & \$91,522.54 & 637,381 & 184,919 \\
\hline
\end{tabular}
\caption {Summary of Asylum Stresser and Lizard Stesser leaked databases and scraped VDO reported data. \(^\dagger\) Revenue was converted from bitcoin to USD.
*Revenue is estimated based on subscription cost and number of paying subscribers. \(^\ddagger\) Domain name is abbreviated to the first three characters.}
\label{tab:scrape_leak}
\end{table*}

\section{Related Work}

DDoS attack and defense techniques have been studied for close to two decades~\cite{ddos_taxonomies,
reflectors,Traceback,low_rate,pushback,botz,AmplificationHell,TCPAmplification}.
There have also been several empirical studies of DDoS attacks in the wild using backscatter
analysis~\cite{backscatter} which were revisited by
Wustrow \textit{et al.}~\cite{Wustrow10imc}. More recent studies have measured Network Time Protocol (NTP) based DDoS attacks~\cite{ntp}
and conducted broader measurements of UDP amplifiers along with introducing methods to identify spoofing-enabled networks~\cite{ddoshell}.

Other studies have explored the structure of botnet based DDoS attacks~\cite{ddos_botnets} and malware~\cite{ddos_botnets_survey,zombie_roundup}.
However, the closest related work to ours in this vein is by K\"uhrer \textit{et al.} which monitored the impact of DDoS attacks on victims~\cite{DDosBotnets}
and an analysis of a leaked database from a single booter service done by Karami and McCoy~\cite{booter}.
Our work differs from this previous work in that we are focused on holistically understanding the stakeholders and infrastructure these booter
services rely on to operate across a larger set of booter services.

Our study is in the same vein as prior work that views security problems through an economic lens~\cite{Tyler:Economics}.
We set out to understand the stakeholders and infrastructure of criminal DDoS-for-hire enterprises as has been done in other domains,
such as abusive advertising~\cite{trajectory, mccoy2012pharmaleaks, searchSeizure} and fake anti-virus~\cite{fakeAV}.
Since booters are a criminal support service rather than the previously studied domain of abusive advertising, they operate under a
different set of constraints. In this respect our work is more along the lines of studies focused on criminal support services, such as 
email spam delivery~\cite{botnetmaster, kanich2008spamalytics}, fake social links~\cite{green} and fake account creation~\cite{thomas2013trafficking}. While this same
\textit{follow-the-money} and payment intervention approach has been explored in previous studies~\cite{priceless}
our study is the first, to our knowledge, that explores the effectiveness of this method against more robust criminal-to-criminal payment methods rather 
then consumer-to-criminal payments.

The largest contribution of our study is in characterizing the ecosystem of subscription-based booter services, which has not been studied 
in much depth. We show that these booters are structured differently than traditional botnet based DDoS services that are rented for a fixed 
time period in terms of the underlying attack infrastructure, customer base, business model and payment methods.
We believe that our findings enable a better understanding of the effectiveness of ongoing efforts to disrupt their attack infrastructure 
at the amplifier and hosting level, a preliminary evaluation of the potential effectiveness of a payment intervention
along with a detailed analysis of the nature of these services and how they are structured.

%% file: sections/users_victims.tex
\section{Inside View of Booters}
\label{sec:scale}

In this section, we analyze publicly leaked backend booter databases, scraped data and aggregated PayPal account data. From this
analysis we present some numbers to better understand the dynamics and scale of booter services. This includes, the amount of revenue generated,
the number of users and their probable geolocation, the number of victims and the number of attacks initiated by the subscribers of these services.

\subsection{Data Sets}

\begin{table}[t]
\small
\centering
\begin{tabular}{lrlr}
\hline
\multicolumn{2}{c}{Operator} & \multicolumn{2}{c}{Subscriber} \\
\hline
\bfseries CC & \bfseries \% & \bfseries CC & \bfseries \% \\
\hline
US & 44.06\% & US & 47.58\% \\
PK & 15.03\% & DE & 10.45\% \\
CA & 13.99\% & GB & 5.60\% \\
GB & 6.29\%  & NL & 4.87\% \\
AU & 3.15\% & RU & 4.81\% \\
\hline
\end{tabular}
\caption {Top country IP geolocations of logins for booter operator's and subscriber's PayPal accounts.}
\label{operatorAccounts}
\end{table}

Our datasets for this section are comprised of aggregated PayPal account geolocation data provided by PayPal, two leaked 
backend databases for Asylum and Lizard Stresser and scraped data from VDO. A summary of these data sets is included in
Table~\ref{tab:scrape_leak} and Table~\ref{operatorAccounts}. Before presenting our analysis we
will first describe each of these data sets in more detail. 

\noindent\textbf{VDO Scraped Data.} At the time we started monitoring \textit{VDO} to measure the scale of its 
operation in early December 2014, it was one the top booter services on underground forums with a high rate of positive reviews. 
During an eight weeks period ending 
in early February 2015, we crawled this booter every 10-minutes to collect data on users of the service and details of attacks 
launched by them. We found \textit{VDO} to be unique in reporting a wealth of public information on their users and attack 
details. This data includes all users that logged into the services in the past 15 minutes and distinguishes paying subscribers from
unpaid users. In addition, it displays a list of all currently running attacks that includes the type of the attack, the target of the 
attack, duration and the time remaining for the attack to be finished. Users can optionally choose to remain anonymous and hide attacks, 
but the default is for all information to be public. Less than 30\% of login records scraped were anonymous and 39\% of all attacks 
seen were hidden. 

While we cannot fully vet this self-reported data, we did verify that the data representing our actions were reported accurately. 
We also validated that all NTP attacks reported for a day were accurate by sending monlist requests in 10-minute intervals to a set 
of 12 NTP amplifiers known to be abused by \textit{VDO} and recorded the received responses. A total
of 44 distinct victims were the target of NTP attacks as reported by \textit{VDO} during that 24 hour time period and we were able to 
find matching records for all 44 targets in the monlist responses collected from the set of 
monitored NTP servers. This gives us some increased level of confidence that the details of reported attacks and users are accurate.

\noindent\textbf{Asylum Stresser Backend Database.} Asylum Stresser was an established booter that was in operation for over 
two years before their backend database was publicly leaked, which covered 18 months of operational data that included
user registrations, payments and attack logs. It ceased operation shortly after the leak and has not resumed
operation. This leaked database has been vetted by many members of the anti-DDoS community that located their own
test accounts in the user registration data and is believed to be authentic.

\noindent\textbf{Lizard Stresser Backend Database.} Lizard Stresser was launched in late December of 2014 by 
individuals calling themselves the \textit{Lizard Squad}. This same group 
was responsible for DDoS attacks on Sony PlayStation and Microsoft Xbox networks on December 25, 2014. Later, the group announced 
that the attacks were meant to 
demonstrate the power of Lizard Stresser, a booter service they started to offer to users on a subscription basis. As the attack 
infrastructure used by this service was backed by hacked home Internet routers, we did not directly interact with this service. 
However, their backend database covering their first two weeks of operation that included user registrations, payments and attack logs 
was publicly leaked. For this database, since 
all payments were in bitcoin and the wallet addresses are included,
we have validated that this part of the database is accurate. We have also checked for internal consistency within these leaked
databases. While we cannot rule out that some of the data has been fabricated, it would take a fair amount of resources to
create this high fidelity of a forgery.

\noindent\textbf{Aggregated PayPal Data.} In order to understand where booter operators and subscribers are potentially located,
we use aggregated data provided to us by PayPal that was computed from all the accounts identified by PayPal as belonging to booter operators
and subscribers. This data did not include any scale on the number of booter and subscriber accounts included in the dataset. It was
computed by assigning the location for each account to the country from which the majority of their logins occurred and computing the percentage of
accounts assigned to each country. In the case that an account did not have a majority of their logins occurring in a single country
it was removed from the dataset. This accounted for 3\% of subscriber accounts and none of operator accounts. Also, IP addresses
for proxies, VPN services, hosting service and Tor were removed using a database from IP2Location~\footnote{IP2Proxy IP-Country Database---\url{http://www.ip2location.com/databases/ip2proxy}}

\subsection{Subscribers}

We find that 15\% of Asylum users and 23\% of all VDO users purchased a subscription, however less than 2\% of all Lizard Stresser
users purchased a subscription. This might be attributed to the fact that Asylum and VDO both accepted PayPal payments at least
sporadically while Lizard Stresser only accepted Bitcoin as a payment method. It is difficult to attribute why the conversion rate
of registered users to subscribers is much less for Lizard Stresser, since other factors, such as the media coverage, might have also 
driven many users to sign up out of curiosity. However, anecdotally 42 of the 225 support tickets mention that the user wants to
purchase a subscription using PayPal. As one potential attacker wrote, ``I want to pay via paypal real bad I'm a huge fan of and 
want to buy this ASAP but I don't have bitcoins.''

Based on the aggregated geolocation information provided by PayPal in Table~\ref{operatorAccounts}, 
over 44\% of the customer and merchant PayPal accounts associated with Booters are potentially owned by someone in the United
States.
There are many inherent limitations of this data which we cannot correct or quantify due to the highly aggregated nature of the data.
These include the fact that
booter services create multiple accounts to replace ones that are limited by PayPal. Thus it might be the case that a few booter
services control a large number of the total accounts and are biasing the location and the same might hold for customers as well.
Assuming it is accurate, this might support the notion that these services are to some extent limited in their ability to accept
payments using virtual currencies, such as Webmoney, which block residents in the United States from creating accounts.

%the operators of lizardstresser included Brian Kreb's PayPal
%address on their site as a prank. Through private communications Brian Krebs has reported receiving a number of payments that 
%were intended as subscription payments for lizardstresser. This suggests that lizardstresser lost revenue from users that mistakenly 
%attempted to use PayPal to pay for their subscriptions. Larger scale studies of leaked databases and measurements of payment
%method avalibility might be able to quantify how this impacts subscribers and revenue.

%To keep the back-end attack servers unsaturated, very long attacks are not allowed. The average attack duration is 1,669 seconds.  

%\begin{table}[H]
%\small
%\centering
%\begin{tabular}{|l|c|}
%\hline
%\bfseries Attack Type & \bfseries \% \\

%\hline
%SSDP  &  42.75\% \\
%Joomla & 11.90\% \\
%ESSYN  & 11.43\% \\
%NTP  & 9.04\% \\
%DNS & 8.72\% \\

%\hline
%\end{tabular}
%\caption {Distribution of top attack types}
%\end{table}

%As reported by previous research, the targets of attacks are mostly located in the US and western Europe. 

%\hl{While Brian has reported that the booter has earned USD \$11,000 worth of bitcoins during its operations, it's not clear how this number is calculated. There is a bitcoin table with 7260 records and a total sum of 1630 BitCoins. This number obviously does not make sense. Most of the records have an status of 0. However, the status for our own transaction is 10. There are only 181 records with the status code of 10 and a total sum of 12.447 BitCoins.}

\subsection{Revenue}

\begin{figure}[t]
  \centering
  \includegraphics[width=1.00\linewidth]{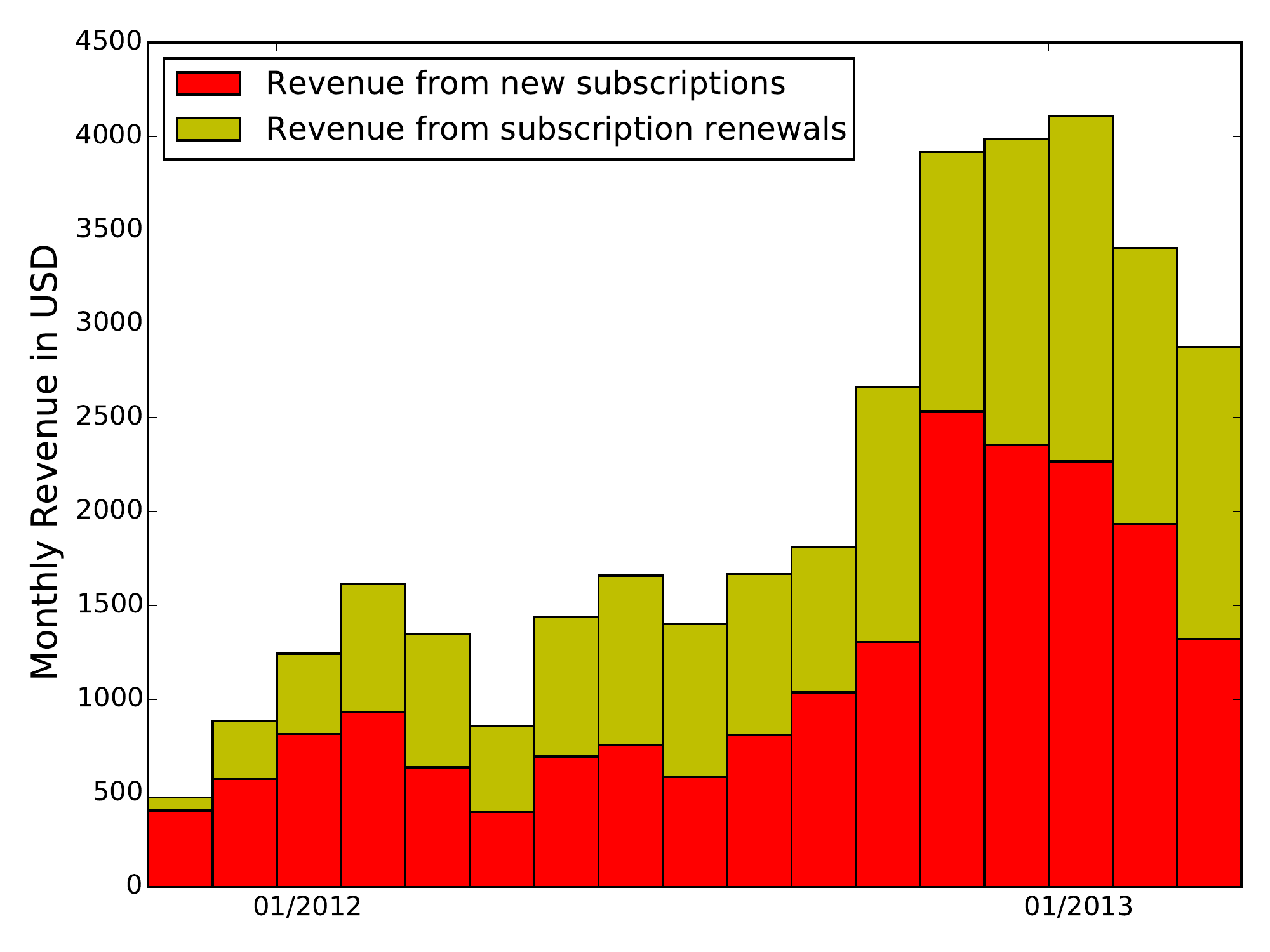}
        \caption{Asylum Stresser monthly revenue.}
        \label{fig:asylum_monthly_revenue}
\end{figure}

Ayslum earned an average of \$2,079/month. However, as Figure~\ref{fig:asylum_monthly_revenue}
shows their revenue started at a modest \$500/month and grew to over \$3,000/month towards the end of the leaked data period. Also, it is interesting to note that
their revenue from subscription renewals was \$16,025.12 and almost equal to the \$19,356.42 earned from new subscriptions.
The Lizard Stresser leak only covers 2 weeks in which time they earned \$3,368 and VDO earned an estimated \$24,737/month confirming that VDO was operating 
at a far larger scale. 
Asylum collect 99.4\% (\$35,180.14) of their revenue through PayPal payments and only 0.6\% (\$201.40) of their revenue was collected using their secondary
payment method of MoneyBookers. Lizard Stresser collected all their revenue through their only supported payment method of Bitcoin and VDO accepted both 
PayPal and Bitcoin. All of this underscores
the fact that revenue from paid subscriptions and renewals is the driving factor for operating these services and expanding them to grow customer bases.
They are presumably profitable, but these individual booters do not generate the profits required to pay the upfront capital, fees and potential fines
for dedicated credit card merchant processing accounts, which amounted to around \$25-\$50K per an account, as was the case with illicit pharmaceutical
and fake anti-virus groups that had revenues on the order of millions of USD dollars a month~\cite{mccoy2012pharmaleaks,fakeAV}. 

\subsection{Attacks}

From the leaked data we find that these three booters were responsible for over half a million separate attacks against over 100,000
distinct IP addresses. While the average attack from VDO only lasted 27 minutes this data demonstrates the large-scale abuse
problems and unwanted traffic generated by these services. Our analysis of victims finds that they are predominantly residential
links and gaming-related servers, with a small number of higher profile victims, such as government, journalist and law enforcement
sites. This matches previous analysis of victims from leaked databases~\cite{booter}. For VDO our scraped data included the
type of DDoS attack launched and our analysis of this data shows that amplified attacks where the adversary attempts to exhaust 
the bandwidth capacity of the victim's connection accounted for 72\% of all attacks launch from VDO. The next most popular class 
of attacks were SYN flooding attacks, which made up only 16\% of all attacks.

%The only source of concern about the transactions was the large number (165) of distinct transaction values. I'm not sure if we need to cut out some values? All numbers are below \$100 and therefore reasonable.
%\href{https://docs.google.com/spreadsheets/d/1A8uRftX73Ohi6trslpTk6AHZBHGfWt_7uUolmAcf6E8/edit?usp=sharing}{Google docs link}

%% file: sections/infrastructure.tex
\section{Attack Infrastructure}
\label{attack}

Our measurements of booters' attack infrastructure are based on engaging with these services to understand 
what techniques and hosts are being actively used for attacks. Using this information
we provide a better understanding of cost structure and trade-offs of different attack techniques. It also 
informs defenders as to which ISPs and hosts to focus on for blacklisting, 
remediation and notification efforts. Our analysis of frontend servers finds a reliance on CloudFlare
to protect this infrastructure from takedown and DDoS. In addition, we find that booters gravitate to using more 
stable amplifier infrastructure when possible. This differs from previous studies that scan the Internet for the 
vulnerable populations of misconfigured amplification servers many of which might be highly transient and not 
be used for DDoS attacks. We also identify two hosting providers connected to the same ISP that are actively
courting booter operators and providing stable high bandwidth attack servers that allow spoofing. 

This set of measurements is based on data gathered from a combination of sources including subscribing to 
booters and launching attacks against our server, active probing measurements and analysis of two hosting 
providers that rent attack servers. 

\subsection{Dataset}

Our first task was to identify booter services for this part of our study. Absent a centralized location for 
finding booters, we found services via search engines and advertisements on hacker forums such as 
\textit{hackedforums.net}. We selected 15 booter services that received the most positive feedback on underground forums
for our attack infrastructure characterization. The number of booters was kept relatively small in order to minimize the 
amount of money we paid to these services, which totaled less than \$140 and no individual booter service received more than \$19. 
We make no claim about the coverage these booters provide of the entire ecosystem. Rather, we were looking to 
provide a sample of stable services most of them ranked highly for search terms associated with booter services
and they also garnered the most positive replies to their advertisements on underground forums.

We purchased a one month subscription from each of the services which ranged from \$2.50-18.99 and
focused on measuring amplification attacks based on our measurements of VDO that showed this was the most common type 
of attack and the fact that amplification attacks were the default attack type for all 15 booters.
More precisely, we chose to measure the most common amplification reflection attack types offered by the booters, which were SSDP, NTP, 
DNS and Chargen. Table~\ref{infrastructure-booters} shows the set of booters, which of the four attack types
that booter offered and the cost of a basic month subscription. 
%We protect our identities from the booter
%services by using multiple PayPal accounts.

We conducted attacks directed at our target server from December 2014 - January 2015. 
The goal of these attacks was to map out the set of misconfigured hosts that
were being used by each booter to amplify their reflection attacks. 
The configuration of our target system used for measuring the attacks was an Intel Xeon 3.3GHz server running Ubuntu 
with 32 GB of RAM and an isolated 1~Gbps dedicated network connection.

We used gulp~\cite{gulp}, which is a lossless Gigabit packet capture tool to capture attack traffic. Each attack lasted
for one hour total and was comprised of many shorter attack instances of 10 minutes each, which is the standard
time limit for basic subscriptions. The reasoning behind the longer attack times was to increase our probability
of identifying all the misconfigured reflection hosts used by a booter for each attack type. However, we found almost all
of the amplification servers were identified after the first attack. Therefore we limited our attacks
to shorter durations for the rest of our experiments that required launching attacks.
%identified amplifiers in the one hour attack were observed in the first attack instance. For NTP we only consider 
%amplifiers responding to spoofed monlist requests.
  
\begin{table}[t]
\footnotesize
\centering
%\resizebox{\linewidth}{!}{%
\begin{tabular}{llc}
\hline
\bfseries Booter & \bfseries Attack Types & \bfseries Cost \\
\hline
ANO & DNS & \$6.60\\
BOO & NTP,Chargen & \$2.50 \\
CRA & DNS,SSDP & \pounds10.99 \\
GRI & NTP,SSDP & \$5.00 \\
HOR & NTP,SSDP & \$6.99  \\
INB & DNS,NTP,SSDP & \$11.99 \\
IPS & NTP,SSDP,Chargen & \$5.00  \\
K-S & SSDP,Chargen & \$3.00 \\
POW & SSDP & \$14.99 \\
QUA & DNS,SSDP & \$10.00 \\
RES & DNS,NTP & \$10.00 \\
SPE & DNS,NTP,SSDP,Chargen & \$12.00  \\
STR & DNS,SSDP & \$3.00 \\
VDO & DNS,NTP,SSDP & \$18.99 \\
XR8 & DNS & \$10.00 \\
\hline
\end{tabular}
%}
\caption {List of booter services we measured, which of the four attack types included in our measurements each booter offered and cost of cheapest one month subscription.}
\label{infrastructure-booters}
\end{table}

\vspace{5mm}  
   
\subsection{Frontend Servers}

Booter services maintain a frontend website that allows customers to purchase subscriptions and launch 
DDoS attacks using convenient drop-down menus to specify the attack type and victim's IP or domain name. 
These frontend websites commonly come under DDoS attack by rival booters and are subject to abuse complaints 
from anti-DDoS working groups. All 15 booters in our study use CloudFlare's DDoS protection services
to cloak the ISP hosting their frontend servers and to protect them from abuse complaints and DDoS attacks.

As part of this study we contacted CloudFlare's abuse email on June 21st 2014 to notify them of the abusive 
nature of these services. As of the time of writing this paper, we have not received any response to our complaints
and they continue to use CloudFlare. 
This supports the notion that at least for our set of booters CloudFlare is a robust solution to protect 
their frontend servers.
In addition, \textit{crimeflare.com} has a list of over 100 booters that are using CloudFlare's services to
protect their frontend servers.

\subsection{Attack Servers}

\begin{table}[H]
\small
\centering
\resizebox{\linewidth}{!}{%
\begin{tabular}{ccccc}
\hline
\bfseries Provider & \bfseries VPS IP & \bfseries Uplink speed & \bfseries Bandwidth & \bfseries Monthly cost\\
\hline

CaVPS Host & 192.210.234.203 & 3.5 Gbps & Unmetered & \$35 \\
Spark Servers & 96.8.114.146 & 949 Mbps & 10 TB & \$60 \\

\hline
\end{tabular}}
\caption {Spoofing enabled VPS services.}
\label{spoofing_services}
\end{table}

\begin{table*}[t]
\small
\centering
%\resizebox{\linewidth}{!}{%
\begin{tabular}{crrrrrrrr}
\hline
\bfseries  & \multicolumn{2}{ c }{\bfseries Chargen} & \multicolumn{2}{ c }{\bfseries DNS }& \multicolumn{2}{ c }{\bfseries NTP} & \multicolumn{2}{ c }{\bfseries SSDP} \\
\hline
\bfseries Booter & (\#) & (\%) & (\#) & (\%) & (\#) & (\%) & (\#) & (\%) \\
\hline
ANO & - & - & 1,827 & 73\% & - & - & -  & -\\
BOO & 370 & 65\% & - & - & 1,764 & 86\% & - & - \\
CRA & - & - & 43,864 & 56\% & - & - & 64,874 & 46\% \\
GRI & - & - & - & - & 1,701 & 72\% & 10,121 & 60\% \\
HOR & - & - & - & - & 8,551 & 58\% & 242,397 & 30\% \\
INB & - & - & 38,872 & 55\% & 4,538 & 92\% & 170,764 & 54\% \\
IPS & 1,636 & 44\% & - & - & 1,669 & 85\% & 90,100 & 29\% \\
K-S & 1,422 & 30\% & - & - & - & - & 5,982 & 76\% \\
POW & - & - & - & - & - & - & 1,424,099 & 11\% \\
QUA & - & - & 10,105 & 85\% & - & - & 39,804 & 67\% \\
RES & - & - & 2,260 & 82\% & 27 & 100\% & - & - \\
SPE & 2,358 & 38\%  & 26,851 & 61\% & 6,309 & 35\% & 258,648 & 24\% \\
STR & - & - & 93,362 & 53\% & - & - & 7,126 & 74\% \\
VDO & - & - & 16,133 & 82\% & 6,325 & 82\% & 150,756 & 62\% \\
XR8 & - & - & 44,976 & 52\% & - & - & - & - \\ \hline
Total & 4,565  & 23.46\% & 181,298  & 35.30\% & 17,599  & 42.31\% & 2,145,015 & 11.84\% \\
\hline
\end{tabular}
%}
\caption {Number of total amplification servers and percentage of overlap with amplifications servers used by other booters.}
\label{totalNumber}
\end{table*}

Renting back-end servers to generate attack traffic is the primary source
of cost for the operators of booter services. We did some research to get a broad sense of the market availability
and cost of back-end servers that allow the source IP address to be spoofed. Being spoof friendly, fast uplink speed and
high caps or unmetered bandwidth usage are the key requirements of a server appropriate for supporting the operation of a booter service.

In order to understand if these services delivered on their claims of allowing spoofing and providing the bandwidth they
advertised we rented Virtual Private Servers (VPS) from two hosting providers that advertised on underground forums.
Table~\ref{spoofing_services} summarizes the services that we purchased.
Both of the VPSs we purchased were connected to the same ISP (ColoCrossing) in the US. We also verified that both VPSs allowed
spoofing and measured their actual link speeds. One VPS provided around 1Gbps uplink bandwidth and the other one interestingly
provided up to 3.5Gbps. The servers we purchased were virtual and therefore running on shared hardware. A busy booter service
would need to use dedicated servers with more resources to support its operation. The price range for a dedicated servers
with a link speed of 1Gbps and unlimited bandwidth usage was around \$300-\$500. Due to budget limitations we could
only rent these two VPSs and did not rent any higher end dedicated servers. However, our initial results show that this is a
potentially effective method of mapping out abusive hosting and we plan to scale this part of our measurements as future work.

\subsection{Attack Techniques}

Due to their effectiveness, amplified volume-based attacks are the default attack technique offered by most
booter services. We focused our analysis on SSDP (more commonly known as Universal Plug and Play (UPnP)), DNS,
NTP and Chargen.
These attacks depend on servers running misconfigured UPnP, DNS resolvers, NTP and Chargen services that
enable attackers to amplify attack traffic by sending spoofed packets with the victim's source address in the IP header
and having these services respond with a larger amount of traffic directed to the victim.

We have also seen booter services offering reflection-based attacks by misusing popular web Content Management Systems
(CMS) such as WordPress and Joomla to generate and direct HTTP requests to target web servers.

In addition, many booters offer direct attacks including TCP SYN, and UPD flood where the attack spoofs the source
IP address and directly sends packets to the victim. Some booters also implement HTTP-based attacks including HTTP
POST/GET/HEAD, RUDY (R-U-Dead-Yet) and Slowloris using proxies to conceal their attack server's IP address.

\subsection{Amplifiers}

As part of our measurements we can map out the set of amplifiers that are being abused to magnify the traffic
volume of attacks.
This sheds light on the population of hosts that are not only vulnerable to amplification attacks, but are actively
being abused. Table~\ref{totalNumber} shows that the set of abused Chargen and NTP servers are smaller and
more highly shared between two or more services, whereas there is an ample supply of DNS and SSDP servers
that are used as amplifiers. However, the overlap of DNS servers used by two or more booter services is still
relatively high suggesting that these DNS resolvers might be more stable, have higher bandwidth connections and
be in more limited supply.

\begin{table}[t]
\small
\centering
\resizebox{\linewidth}{!}{%
\begin{tabular}{cclc}
\hline
\bfseries CC & \bfseries \ \% & \bfseries \ AS & \bfseries \% \\
\hline
\multicolumn{4}{ c }{\bfseries Chargen} \\ \hline
CN & 48.78\% & 4134 (Chinanet) & 14.46\% \\
US & 12.51\% & 37963 (Hangzhou Alibaba Advertising) & 10.47\% \\
KR & 5.50\% & 4837 (CNCGROUP China169 Backbone) & 6.88\% \\
RU & 4.58\% & 17964 (Beijing Dian-Xin-Tong Network) & 2.61\% \\
IN & 2.56\% & 7922 (Comcast Cable Communications) & 2.61\% \\
\hline
\multicolumn{4}{ c }{\bfseries DNS} \\ \hline
US & 12.38\% & 4134 (Chinanet) & 2.68\% \\
RU & 11.58\% & 3462 (Data Communication Business Group) & 2.15\% \\
BR & 9.19\% & 18881 (Global Village Telecom) & 1.46\% \\
CN & 6.84\% & 4837 (CNCGROUP China169 Backbone) & 1.45\% \\
JP & 3.61\% & 7922 (Comcast Cable Communications) & 1.27\% \\
%TW & 3.08\% & 9121 (Turk Telekomunikasyon Anonim Sirketi) & 1.22\% \\
\hline
\multicolumn{4}{ c }{\bfseries NTP} \\ \hline
US & 31.47\% & 3462 (Data Communication Business Group) & 14.01\% \\
TW & 15.29\% & 46690 (Southern New England Telephone) & 12.35\% \\
CN & 10.68\% & 7018 (AT\&T Services) & 4.84\% \\
KR & 5.50\% & 4134 (Chinanet) & 3.58\% \\
RU & 4.74\% & 4837 (CNCGROUP China169 Backbone) & 2.18\% \\
\hline
\multicolumn{4}{ c }{\bfseries SSDP} \\ \hline
CN & 36.26\% & 4837 (CNCGROUP China169 Backbone) & 18.98\% \\
US & 19.37\% & 4134 (Chinanet) & 11.16\% \\
EG & 6.83\% & 8452 (TE Data) & 6.61\% \\
AR & 5.37\% & 22927 (Telefonica de Argentina) & 5.13\% \\
CA & 5.36\% & 7922 (Comcast Cable Communications) & 4.60\% \\
\hline
\end{tabular}
}
\caption {Top country locations and autonomous systems for amplifiers.}
\label{ccas}
\end{table}

\subsection{Amplifier Location}

As demonstrated by Table~\ref{ccas} both the geolocation and AS of amplifiers used by booters are fairly diffuse.
There are a few notable exceptions, such as the concentration of Chargen amplifiers in China with three Chinese
ASs connecting 34\% of these amplifiers. In addition, there is a slight concentration of abused NTP servers
connected to one Taiwanese AS and two United States network operators. This might indicate a potential to focus
notification and patching efforts on these networks, given the limited pool of hosts used for Chargen
and NTP attacks from Table~\ref{totalNumber}. Feeds of these actively abused servers could also be distributed to these
network operators and to DDoS mitigation services.

\begin{figure}[t]
  \centering
        \includegraphics[trim = 20mm 0mm 25mm 10mm, clip, ,width=1.00\linewidth]{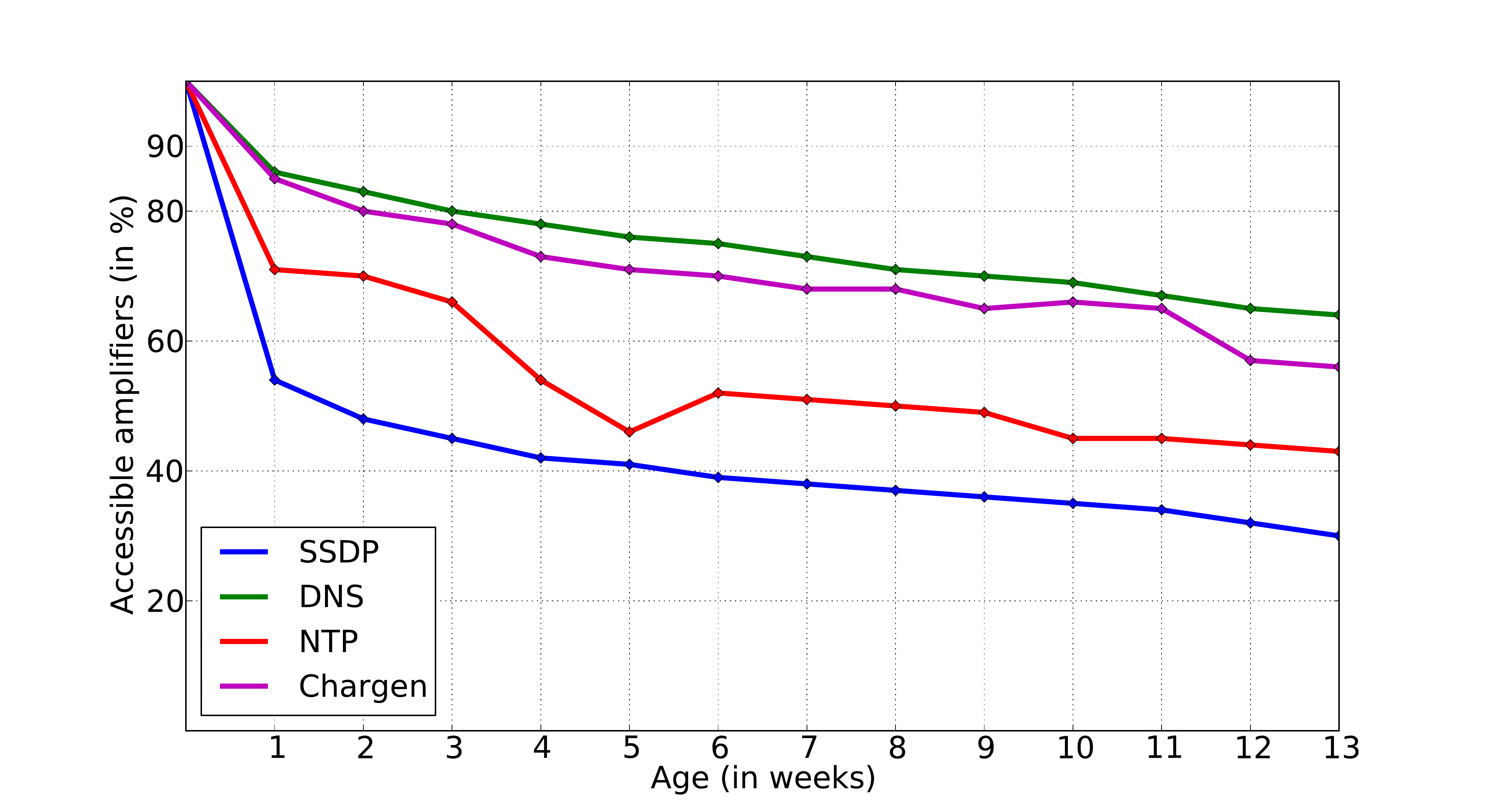}
        \caption{IP churn of amplifiers.}
        \label{fig:pp_account_lt}
\end{figure}

\subsection{Amplifiers Churn}

In order to measure the stability of these amplifiers we probed them periodically for 13 weeks to understand how many were still
located at the same IP and vulnerable to abuse. As shown in Figure~\ref{fig:pp_account_lt}, the set of DNS resolvers 
were the most stable with nearly 80\% still vulnerable and located at the same IP after one month and over 60\% were still misconfigured
after 13 weeks. 
This result is counter to the previous results of churn based on Internet wide scanning that found a 50-60\% churn rate for DNS 
servers after one week~\cite{ddoshell}. It indicates that booters have gravitated to using a more stable set of DNS resolvers
and that focusing mitigation efforts on these might cause these DNS attacks to be less efficient and require additional bandwidth and
cost. From our measurements, SSDP servers where the least stable with a 50\% churn rate after only a single week. This result agrees 
with the previous Internet wide scanning result and indicates that either booters have not found or there might not exist a set of 
more stable SSDP servers.

\subsection{Amplification Factor}

One of the few direct costs incurred for every attack a booter service launches is the bandwidth sent from their rented
attack servers. In order to reduce this cost amplification attacks are used for volume-based flooding attacks.
Some attack methods can potentially produce a larger amplification factor then others, but there are many factors
that effect the amplification factor. Our measurements largely agree with the lower-end bandwidth amplification factor 
numbers reported in a previous study~\cite{ddoshell}, with NTP attacks resulting in an average amplification factor of
603 times, Chargen being the next largest at 63 times, DNS resulting in an average of 30 times amplification and SSDP generating the smallest
average amplification factor of 26 times. This and the limited number of NTP amplifiers confirms that the communities
focus on prioritizing notification and patching of misconfigured NTP servers is the correct approach. We also suggest that 
some effort be placed into efforts to notify operators of servers with misconfigured and abused Chargen services, since these
are the next largest threat and there is also a potentially constrained supply.

%% file: sections/paypal_intervention.tex
\section{Payment Intervention}

As part of our study we sought out opportunities to understand and also measure the effectiveness of 
undermining DDoS Services. In this section, we present our measurements of a payment intervention
that was conducted in collaboration with PayPal.

We find that reporting accounts to responsive payment service providers, such as PayPal, can have the desired 
effect of limiting their ability and increasing the risk of accepting payments using 
these payment services. Although, this technique requires constant monitoring of the booters and drives 
them to move to more robust payment methods, such as Bitcoin.

\subsection{Payment Ecosystem}

At the onset of our study the majority of booter services accepted credit card payments via PayPal as their primary mechanism for 
receiving funds from their customers. In addition to PayPal, some booters accepted 
bitcoin payments and a limited number of booters also accepted credit card payments using Google Wallet\footnote{Google phased out 
their digital goods payment processing at the start of March 2015 --- \url{https://support.google.com/wallet/business/answer/6107573}.} and virtual 
currencies, such as WebMoney and Perfect Money---these last two prohibit customers from the United States from opening an account and using their platform.

We identified a larger set of 60 booter services~\footnote{This set is larger than the previous set, since we did not have to pay for a 
subscription in order to monitor their payment accounts.} that accepted PayPal and created custom crawlers to monitor their payment 
methods and merchant accounts for about 6 weeks from April 22, 2014 through June 07, 2014. These booters were located from 
underground forum advertisements and web searches for terms commonly associated with booter services.
Again we make no claim about the coverage these booters provide of the entire ecosystem.
To minimize the effect of unstable booters on our study, the final set of booters included in our analysis was limited to the 23 stable booter services that were able to 
successfully use PayPal to receive funds for at least half of the time before the PayPal intervention and used at least one PayPal account after the intervention.  
Nine of these 23 booter services overlapped with the set of 15 booters measured in section~\ref{attack}, among the reasons that six booters
were not included in this measurement is that some did not accept PayPal and others did not accept PayPal over half the 
time in the first 6 week period.

After collecting our initial data on the stability of their PayPal merchant accounts, we reported these booter's domains and accounts to PayPal
and they began to monitor merchant accounts linked to these domains and suspending them after an investigation. Note that PayPal will 
initially limit reported merchant accounts that are found to violate their terms of service by accepting
payments for abusive services and perform an investigation of the account.
Once an account is limited the merchant cannot withdraw or spend any 
of the funds in their account. This will result in the loss of funds in these accounts at the
time of freezing and potentially additional losses due to opportunity cost while establishing a new account.
In addition, PayPal performed their own investigation to identify additional booter domains and limited accounts linked to these domains as well.
This had the affect of a large-scale PayPal payment disruption for the majority of booter services.

In order to further understand the effectiveness of our payment intervention, we monitored underground forums where these booters 
advertise their services and news feeds from booters we joined to discover qualitative data on the effectiveness of PayPal's 
payment intervention.

\begin{table}[t]
\small
\centering
\begin{tabular}{lccc}
\hline
\bfseries Booter & \bfseries accounts before & \bfseries accounts after & \bfseries Status \\
\hline
ANO \(^\star\) & 6 (8.2) & 7 (2.9) & \cmark \\
AUR & 6 (7.2) & 6 (2.7) & \xmark \\
BOO \(^\bullet\) & 6 (8.3) & 11 (2.7) & \cmark \\
CRI \(^\dagger\) & 4 (9.0) & 1 (2.0) & \xmark \\
DAR \(^\diamond\) & 4 (6.0) & 5 (4.8) & \cmark \\
DIA \(^\dagger\) & 3 (15.7) & 0 (-) & \xmark \\
GET & 2 (14.0) & 1 (4.0) & \cmark \\
GRI  \(^\star\) & 4 (10.5) & 1 (6.0) & \cmark \\
HAZ \(^\diamond\) & 4 (12.2) & 5 (5.6) & \cmark \\
IDD \(^\diamond\) & 3 (7.7) & 2 (9.0) & \cmark \\
IPS \(^\diamond\) & 3 (7.3) & 5 (5.4) & \cmark \\
POW & 5 (4.5) & 9 (5.0) & \cmark \\
PRI & 6 (8.8) & 2 (1.0) & \xmark \\
QUA & 11 (4.3) & 22 (1.8) & \cmark \\
RAG \(^\bullet\) & 13 (3.9) & 4 (2.0) & \cmark \\
REB & 2 (11.5) & 9 (2.3) & \xmark \\
RES \(^\star\) & 6 (8.2) & 7 (2.9) & \cmark \\
SNO & 1 (-) & 0 (-) & \xmark \\
STA & 5 (13.0) & 3 (5.3) & \cmark \\
STR &  1 (47.0) & 1 (4.0) & \cmark \\
TIT \(^\diamond\) & 12 (5.3) & 17 (2.9) & \cmark \\
XR8 & 4 (10.5) & 11 (1.6) & \cmark \\
XRS \(^\bullet\) & 8 (5.2) & 4 (2.5) & \xmark \\
\hline
& 119 (7.84) & 133 (3.07) & \\
\hline
\end{tabular}
\caption {Number of PayPal accounts used by monitored booters before and after the intervention. The
numbers within the () are the average lifespan of the accounts used by that booter. Accounts that are
active both before and after are counted only in the before and not included
when computing the average lifespan. Matching symbols
indicate that this set of booters shared at least one PayPal account. These shared accounts might be instances
of a third party agreeing to accept payments for these services.}
\label{pp_account_lt}
\end{table}

\subsection{Usage Pattern of PayPal Accounts}

\begin{figure}[t]
  \centering
        \includegraphics[width=0.50\textwidth]{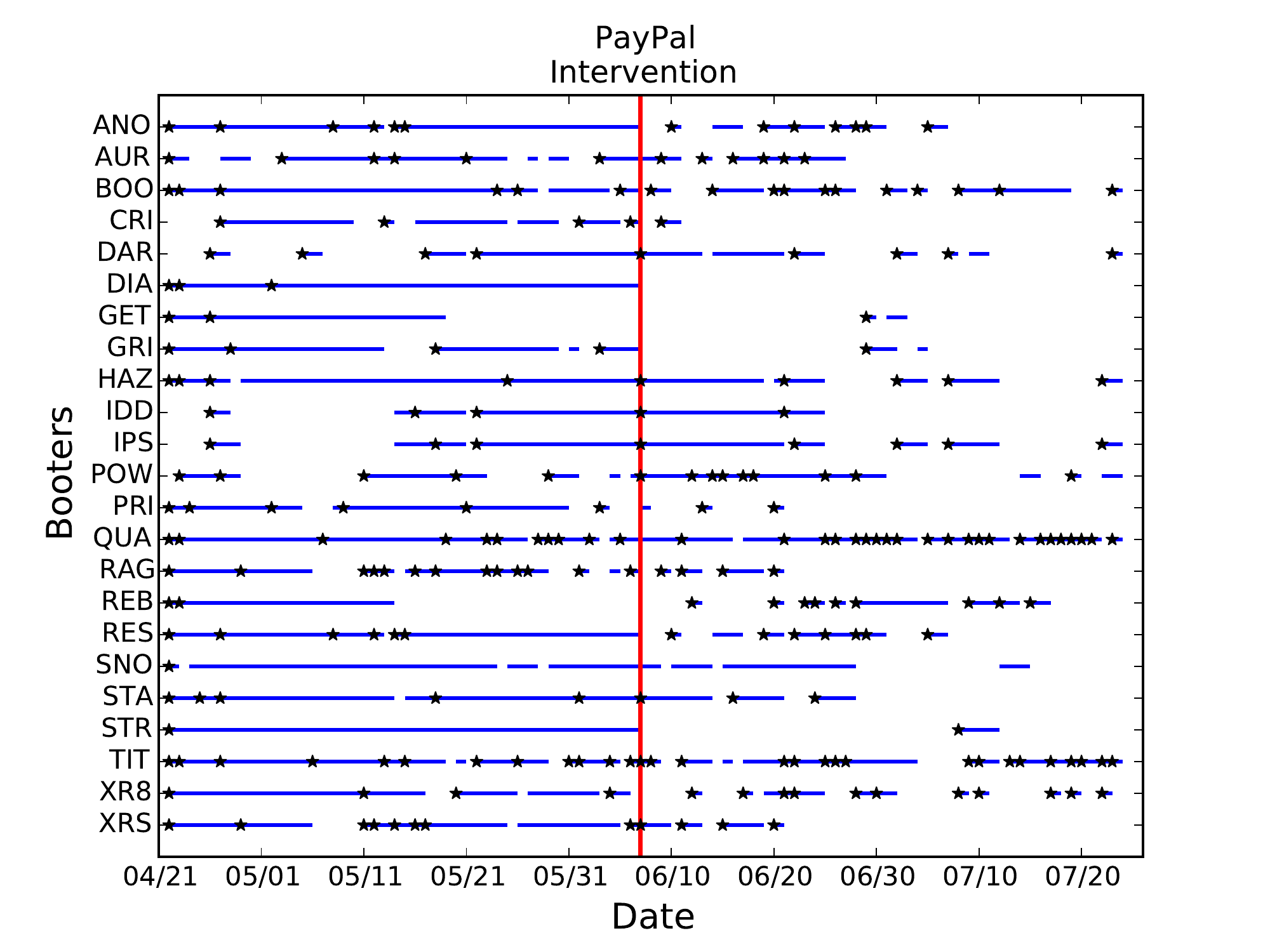}
        \caption{PayPal account usage over time. Black asterisks denote a new PayPal account and gaps in the blue line represent PayPal unavailability for that
time period. The red vertical line indicates when the reporting of accounts started.}
        \label{fig:booter_structure}
\end{figure}

Based on our observations, booter services for the most part only use a single PayPal account at a time to receive payments and change their PayPal 
merchant account when a limit is put on their previous account or they proactively change accounts 
to reduce the risk of limits on their previous accounts. We used the dataset collected during the initial monitoring period to 
understand how frequently booter services were changing their PayPal accounts. Note that our age measures are both right 
and left-censored. For the booter's initial account our data is left-censored and for the last account our data is
right-censored. However, we believe our age measurements accurately represent the effects of the PayPal intervention based
on our interactions with and postings from the booters themselves.

Table~\ref{pp_account_lt} provides an overview of the PayPal accounts observed by our crawler broken down by each service monitored.
As Table~\ref{pp_account_lt} shows accounts had an average life time of slightly over a week before the intervention with
STR and SNO each using a single account that remained active during the entire 47 day initial observation period and SNO's
account that remained active for 37 days after the intervention began.
On the other end of the spectrum, QUA, RAG and TIT changed accounts every 4-5 days before the intervention. The impact of the intervention
can be visually seen in Figure~\ref{fig:booter_structure}.

Once the target intervention begins the average lifespan of an account drops to around 3.5 days with many booter's PayPal accounts
only averaging around two days before they are no longer used again. Figure~\ref{fig:booter_structure} visually shows the impact of the payment 
intervention on the lifespan of 
booter services PayPal accounts and provides some indication of the time period that elapsed between a new PayPal account being actively
used to accept payments and when PayPal took action against the account or it was proactively replaced. The length and number of PayPal outages 
increase after the intervention, with only QUA and TIT avoiding major PayPal outages by resorting to aggressively replacing accounts. Note that this replacement strategy was
not fully effective, since our monitoring infrastructure detected and reported these accounts.

%\begin{figure}[h]
%  \centering
%        \includegraphics[width=1.00\linewidth]{./figures/pp_account_lt.pdf}        
%        \caption{Life time of PayPal accounts used by booter services.}
%        \label{fig:pp_account_lt}
%\end{figure}

\subsection{Booters' Status}

As part of our daily monitoring of the 23 booter services, we recorded if the service could
accept PayPal payments and if the site was functional. This enabled us to better understand 
the impact of the payment intervention on the booter's ability to accept PayPal payments and 
the operation of the service. For each booter, we placed it in one of the following statuses 
each day based on the results of our crawl.
 
\noindent \textbf{Active:} The booter is able to successfully use a PayPal account to receive 
payments from its customers.

\noindent \textbf{Unreachable/Broken:} Either the booter's frontend website was not responding
to HTTP requests, the booter service had closed or the frontend site was not functional.

\noindent \textbf{PayPal Disabled:} The booter's frontend website is active, but the service has 
either removed PayPal as a payment option, or the PayPal account linked to the booter website is limited 
and therefore unable to receive payments.

\begin{figure}[t]
  \centering
        \includegraphics[width=1.00\linewidth]{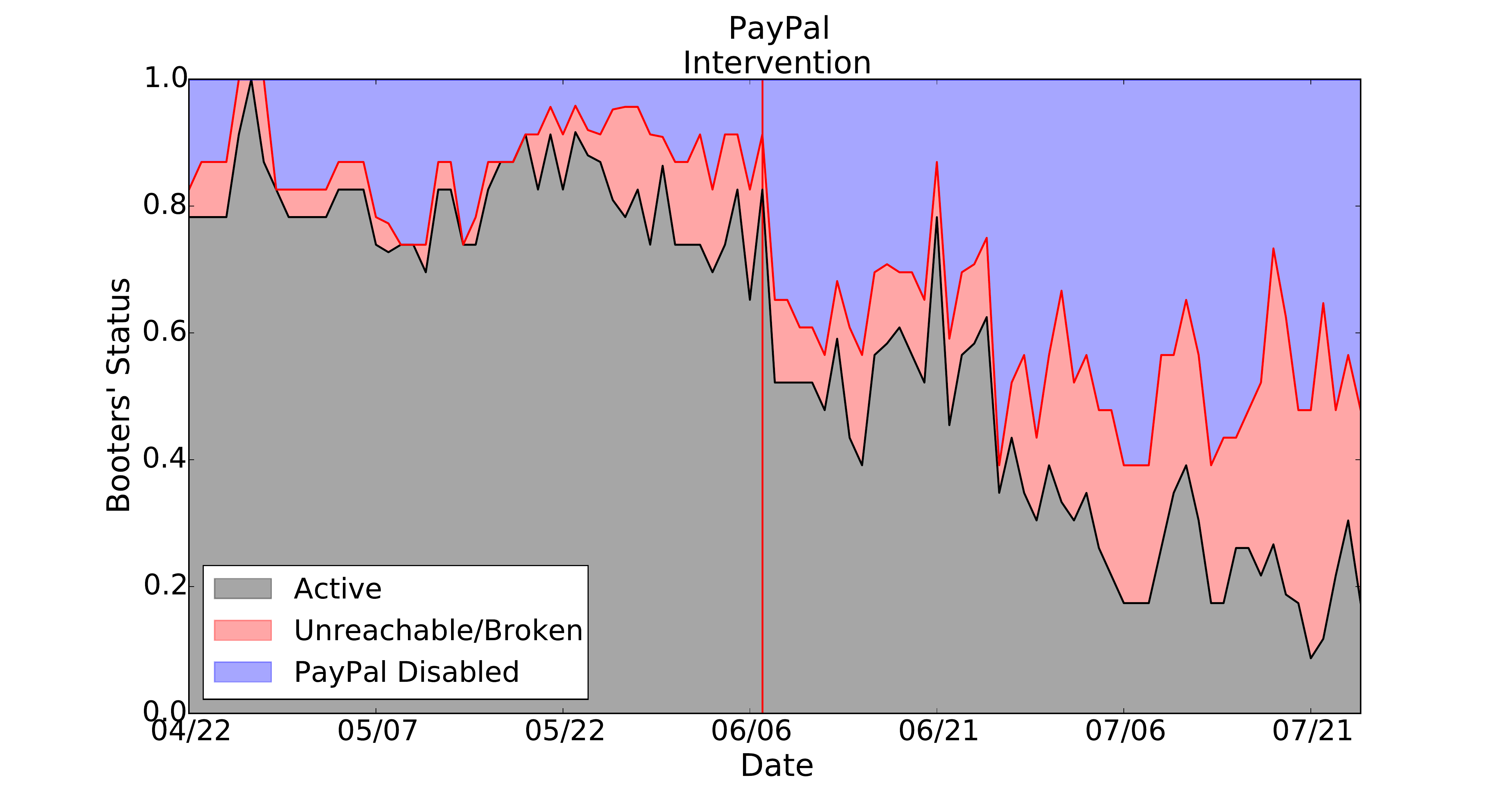}
        \caption{Status of booters over time.}
        \label{fig:pp_booter_status}
\end{figure}

Figure~\ref{fig:pp_booter_status} shows the status of booter services over time. The vertical line 
represents the date on which we started sharing our data with PayPal and PayPal started to 
independently investigate the reported accounts and take action against them. As observed in 
Figure~\ref{fig:pp_booter_status}, the percentage of active booters quickly drops from 70-80\% to 
around 50\% within a day or two following the intervention date and continues to decrease to a low
of around 10\% and then fluctuating between 10-30\%. This resulted in an increase in PayPal
unavailability from 20\% before the intervention to 63\% during the intervention.
In addition, we observed 7 booter services in our study shut down their business 
and most of the remaining services switch to alternative payment methods, such as 
Bitcoin.

%However, as we will show in the next subsection, by switching to 
%less convenient payment methods, such as Bitcoin, the booter services experienced a drop in attacks
%that likely corresponds to a decrease in subscribers and revenue.  

\subsection{Qualitative Assessments}

%From our booter status monitoring we observe that the effect of
%the payment intervention is more dramatic for less established
%booters. We believe that one reason for this is that they have
% not built up enough of a revenue stream and do not have enough
%reserve capital to weather the losses caused by merchant account 
%terminations. This drove many of the smaller booters to shutdown
%their services as shown in the increase of unavailable services
%as the intervention continues.

In addition to our quantitative measurements, we also have qualitative 
evidence of PayPal's payment interventions 
efficacy. By monitoring the underground forums where these services
advertise we can witness the impact of these account limitations.
Wrote one booter operator during the intervention, ``So until now 5 
time my 5 PayPal Accounts got Limited on My stresser is other 
stresser have same Problem with the fucking Paypal ? is there any 
solution what we should do about fucking Paypal ?'' Similarly,
customers vented their frustration at being unable to purchase
a booter service using PayPal. Wrote one booter customer, ``when i 
go to buy a booter it normally says i can't buy because their PayPal 
has a problem.''

In a number of cases booters directly link their closers to loss of
funds due to PayPal merchant account limitations.
This message was posted on the front page of a defunct booter service, 
``It's a shame PayPal had to shut us down several times causing us 
to take money out of our own pocket to purchase servers, hosting, 
and more.''

\subsection{Booter Response}

As with any intervention the adversary, booters, will respond by adapting to
the pressure. In this case, we do not have enough quantitative measurements
to access the effectiveness or the full range of responses to our attempt
to undermine their payment infrastructure. However, we have identified
several common classes of adaptations in response to the intervention.

\noindent\textbf{Alternate payment methods.} Most booters have added
Bitcoin as an alternate payment methods and have posted links
to services that allow customers to purchase bitcoins using 
credit cards or PayPal.
In addition to bitcoins some switched 
to Google Wallet and others added the option to pay using virtual 
currencies, such as Webmoney and Perfect Money. By all accounts
these have resulted in reduced customer bases if the booter cannot 
directly accept credit card payments. Evidenced by the fact
that many booters continued to replace their PayPal accounts
even when previous ones were limited and their funds were lost.
Assuming that alternative payment methods did not result in reduced 
revenue or higher costs there would be little incentive to continue using risky
payment methods, such as PayPal.

\noindent\textbf{Referrer anonymizing services.} We have noticed that
some booters have stopped directly linking to PayPal and are now linking
to an intermediary site and then redirecting the customer's browser
from this intermediary domain to PayPal's site. This intermediary
redirection site is used to hide the booter's real domain name in
the referrer field from PayPal. A subset of booters have also started to replace
this intermediary domain every time they replace a PayPal account.
The effect of this is that it requires active crawling and measurements
of booter sites to identify a booter's new PayPal account 
bypasses passive methods PayPal could use to linking accounts, such as by 
using referrers. This has increased the difficulty of monitoring booter's 
PayPal accounts and effort required to investigate these accounts.

\noindent\textbf{Offline payment.} Finally, in some cases booters have posted
that customers must open a ticket to pay using PayPal. This 
method increases the effort to monitor the booter for new accounts, 
since instead of an automated crawler someone must now interact with
the booter service manually.
It also increases the difficulty of PayPal's investigation into the 
nature of the merchant account. However, this method also requires
the booter service to manually activate each account and drives
away customers that are seeking automated subscription purchasing
systems.

%% file: sections/discussion.tex
\section{Discussion and Future Work}

We have gathered a few key points from ours and the community's efforts to understand and undermine these DDoS services.
Most of these potential strategies involve driving up costs of operating booter services and reducing the convenience of
subscribing. This might force these booters from operating largely unopposed in the open to more
resilient hosting, attack and payment infrastructure for which they pay a premium due to the risk of support
services being taken down or blacklisted for being associated with DDoS attackers.

\noindent\textbf{Reducing scale.} Limiting access to convenient payment methods, such as PayPal, had an impact
on the scale of booter services based on our quantitative and qualitative analysis. However, criminal-to-criminal
payment infrastructure is more resilient then consumer-to-criminal payment infrastructure.
As future work, we plan to understand how to improve the effectiveness of these interventions and make them
sustainable. This in part requires developing more robust monitoring tools that better mitigate countermeasures 
being deployed to make their payment methods more robust to interventions. To this end, we are working on building
a generic crawler that can learn how to navigate a booter site and map out payment accounts. This would reduce
the burden of having to create and maintain custom crawlers for each site.

\noindent\textbf{Reducing effectiveness of attacks.} We plan to continue our monitoring efforts of the amplification servers used
by booters and begin sharing this information with existing patching efforts, such as the OpenResolverProject~\cite{OpenResolverProject}
 and OpenNTPProject~\cite{OpenNTPProject}. Along with this, we plan to experiment with active notifications sent to the ISP and abuse
contact for the server. Our hope is that by focusing mitigation efforts on actively abused amplifiers, we can mitigate the pool of
more stable amplifiers and thus reduce the effectiveness of booter attacks as they are forced to use less stable amplifiers and protocols
with a lower bandwidth amplification factor.
There is some indications that active notification improves patching rates in another context~\cite{heartbleed}.

\noindent\textbf{Increasing costs.} This might be achieved with an increased effort to locate and blacklist or de-peer low-cost 
hosting services that cater to DDoS attacks by providing the ability to spoof and unlimited bandwidth. This might force these services to pay a 
premium for bullet-proof hosting attack servers, which would result in reduced profitability or be passed along to subscribers in the form 
of increased subscription costs. In addition, convincing CloudFlare and other free anti-DDoS services to prohibit these booter 
services would increase their costs by forcing them to build and pay for anti-DDoS services that cater to these abusive booters. 
Admittedly these suggestions will likely not result in large cost increases unless tremendous amounts of pressure were placed
on these parts of their infrastructure.

%\noindent\textbf{Increasing risk to operators.} Our analysis of data provided by PayPal suggests that much of this activity is occurring in the United States.
%If this is the case there is the potential that increased law enforcement efforts could have a direct impact in arresting key operators of these services
%and increasing the perceived risk of operating and using these services. In the case of operators it is likely they could be replaced by overseas
%operators. However, in the case of customers it might be difficult to find a new subscriber based for these services that is located outside the United States
%and Western Europe if the perceived risk of using these services increased. To this end we plan to work with law enforcement to understand how
%effective this type of intervention is on mitigating the threat of booter services.

%% file: sections/conclusion.tex
\section{Conclusion}
Unfortunately, there is no silver bullet that will mitigate the threat posed by booter services 
overnight. These booters have grown in scale due to the perceived low-risk nature, their profitability and 
increasing demand for DDoS attacks as a method of knocking out the competition, harassment and 
censorship on the Internet. 

In this paper we have mapped out a range of support infrastructure that booters depend on in terms of advertising,
attack, hosting and payment. We have also measured the effectiveness of ongoing 
attack infrastructure interventions and demonstrated the potential effectiveness of a payment intervention. 
Our measurement techniques, including direct observations by interacting 
with these booters and support services, such as hosting providers, highlight potential improvements to ongoing 
efforts to undermine attack infrastructure 
by focus patching efforts on abused NTP and Chargen servers and exposing ISPs and hosting centers that enable these attacks. 
We also demonstrated that payment interventions, 
which undermine the accessibility of convenient payment methods, such as PayPal, can potentially have an impact 
on reducing the scale of these services. Our hope 
is that by continuing to explore new methods for understanding and undermining booters, we can identify 
increasingly effective methods of adding friction, cost and risk to these ventures that further erodes 
their attack potency, scale and profitability over time. 

%% file: sections/acknowledgments.tex
%commented out for submission
\section{Acknowledgments}

We thank PayPal's Information Security team for their assistance. This work was supported by National Science Foundation grant NSF-1237076 and gifts from Google.